\RequirePackage{amsthm}

\documentclass[pdflatex, referee, sn-mathphys-num]{sn-jnl}% Math and Physical Sciences Numbered Reference Style 
\usepackage{graphicx}%
\usepackage{multirow}%
\usepackage{amsmath,amssymb,amsfonts}%
\usepackage{amsthm}%
\usepackage{mathrsfs}%
\usepackage[title]{appendix}%
\usepackage[dvipsnames]{xcolor}
\usepackage{textcomp}%
\usepackage{manyfoot}%
\usepackage{booktabs}%
\usepackage{algorithm}%
\usepackage{algpseudocode}%
\usepackage{listings}%
\usepackage{lmodern}
\usepackage{upgreek}
\usepackage[utf8]{inputenc}
\usepackage{fancyhdr} % For customizing headers/footers
\usepackage{xr}
\usepackage{geometry} % For overall page layout
\begin{document}
%% as per the requirement new theorem styles can be included as shown below

\newtheorem{theorem}{Theorem}%  meant for continuous numbers
%%\newtheorem{theorem}{Theorem}[section]% meant for sectionwise numbers
%% optional argument [theorem] produces theorem numbering sequence instead of independent numbers for Proposition
\newtheorem{proposition}[theorem]{Proposition}% 

\theoremstyle{thmstyletwo}%
\newtheorem{example}{Example}%
\newtheorem{remark}{Remark}%

\theoremstyle{thmstylethree}%
\newtheorem{definition}{Definition}%

\raggedbottom
%%\unnumbered% uncomment this for unnumbered level heads

\title[Article Title]{Anticipating Decoherence for Enhancing Coherence in Quantum Systems}

%%=============================================================%%
%% GivenName	-> \fnm{Joergen W.}
%% Particle	-> \spfx{van der} -> surname prefix
%% FamilyName	-> \sur{Ploeg}
%% Suffix	-> \sfx{IV}
%% \author*[1,2]{\fnm{Joergen W.} \spfx{van der} \sur{Ploeg} 
%%  \sfx{IV}}\email{iauthor@gmail.com}
%%=============================================================%%

\author[1,3,7]{\fnm{Pranshu} \sur{Maan}}\email{pmaan@purdue.edu, maan@umich.edu}

\author[1,3,7]{\fnm{Yuheng} \sur{Chen}}\email{chen4114@purdue.edu}

\author[1,2]{\fnm{Sean} \sur{Borneman}}\email{sbornema@andrew.cmu.edu}

\author[5]{\fnm{Benjamin} \sur{Lawrie}}\email{lawriebj@ornl.gov}
\author[4]{\fnm{Alexander} \sur{Puretzky}}\email{puretzkya@ornl.gov}
\author[1,6,7]{\fnm{Hadiseh} \sur{Alaeian}}\email{halaeian@purdue.edu}
\author[1,3,7]{\fnm{Alexandra} \sur{Boltasseva}}\email{aeb@purdue.edu}
\author*[1,3,6,7]{\fnm{Vladimir M.} \sur{Shalaev}}\email{shalaev@purdue.edu}
\author*[1,7]{\fnm{Alexander V.} \sur{Kildishev}}\email{kildishev@purdue.edu}

\affil[1]{\orgdiv{Elmore Family School of Electrical and Computer Engineering, Birck Nanotechnology Center}, \orgname{Purdue University}, \city{West Lafayette}, \postcode{47907}, \state{IN}, \country{USA}}

\affil[2]{\orgname{now at Carnegie Mellon University}, \orgaddress{\street{7325 Wean Hall, 5000 Forbes Ave}, \city{Pittsburgh}, \postcode{15213}, \state{PA}, \country{USA}}}

\affil[3]{\orgdiv{Quantum Science Center (QSC)}, \orgname{Oak Ridge National Laboratory}, \orgaddress{\city{Oak Ridge}, \postcode{37830}, \state{TN}, \country{USA}}}

\affil[4]{\orgdiv{Center for Nanophase Materials
Sciences}, \orgname{Oak Ridge National Laboratory}, \orgaddress{\city{Oak Ridge}, \postcode{37830}, \state{TN}, \country{USA}}}

\affil[5]{\orgdiv{Materials Science and Technology Division}, \orgname{Oak Ridge National Laboratory}, \orgaddress{\city{Oak Ridge}, \postcode{37830}, \state{TN}, \country{USA}}}

\affil[6]{\orgdiv{Department of Physics and Astronomy, Birck Nanotechnology Center}, \orgname{Purdue University}, \city{West Lafayette}, \postcode{47907}, \state{IN}, \country{USA}}

\affil[7]{\orgdiv{ Purdue Quantum Science and Engineering Institute}, \orgname{Purdue University}, \city{West Lafayette}, \postcode{47907}, \state{IN}, \country{USA}}
%%==================================%%
%% Sample for unstructured abstract %%
%%==================================%%

\abstract{Large-scale quantum technologies require coherence across distant nodes, necessitating indistinguishable quantum states. However, environmental disorders, including dephasing, spectral diffusion, and spin-bath interactions, undermine coherence. Using statistical methods, we uncover correlations in decoherence channels induced by slowly varying environments. Spectral diffusion serves as a representative demonstration case that can be extended to other remote, disordered systems such as spins in nitrogen-vacancy centers and quantum-dot spin qubits, as well as flux noise in
superconducting qubits. In this work, we present the first experimental demonstration of internal prediction of unseen spectral dynamics that generalizes across multiple quantum systems, showing that this framework, if implemented, could reduce spectral shift by average factors of approximately 2.32 to 20.19, depending on emitter stability, thereby enabling enhanced coherence and multi-node synchronization for scalable quantum communication, computation, imaging, and sensing.}


%%\pacs[JEL Classification]{D8, H51}

%%\pacs[MSC Classification]{35A01, 65L10, 65L12, 65L20, 65L70}

\maketitle
\section{Introduction}\label{sec1}
Quantum technologies promise breakthroughs in solving classically intractable problems~\cite{Barthe2025, Shor1997}, sensing below the standard quantum limit~\cite{Matsuzaki2018, Hecht2025, Englund2019, Lukin2020, LIGO2024, Mitchell2004, Resch2007, Lawrie2019, Ahmed2018}, and enabling secure communication~\cite{Pittaluga2025, Gisin2007, Bennet2014, Eckert1991, Tittel2002, Brassard2003, Zbiden2021, Scarani2009}. Realizing these promises requires scalable quantum architectures, particularly in the context of quantum networks and distributed quantum systems~\cite{kimble2008, humphreys2018, aghaeeRad2025, main2025, Hanson2018, pompili2021}. Various platforms are being explored, including trapped ions~\cite{Kielpinski2002, Duan2010, Blatt2008, Häffner2005}, neutral atoms~\cite{Evered2023, Covey2023, Lukin2025, Graham2022}, and superconducting circuits~\cite{Wallraff2004, Makihara2025, Arute2019}. However, decoherence remains a central challenge for the realization of scalable quantum technologies, affecting a wide range of platforms including superconducting qubits~\cite{Simon2016}, spin-based systems~\cite{Sanchez2018}, solid-state emitters~\cite{Kim2015}, and trapped-ion architectures~\cite{Glikin2025,Teller2021}, among others.

\noindent Reactive feedback-based schemes attempt to mitigate decoherence by continuously monitoring the optical transition between ground and excited electronic states. The compensatory electric (Stark) or strain fields are then applied to stabilize the emission~\cite{Sohn2019, acosta2012, Hansen2010, Kuruma2025}. However, these techniques are fundamentally limited by feedback latency, the finite bandwidth of the waveform generator, and the measurement back-action, i.e., the disturbance introduced to the quantum system by the act of measurement itself. Since local environmentally induced disorders occur on timescales of nanoseconds to hundreds of milliseconds~\cite{Dephasing2016, santori2004, acosta2012}, even short delays in detection and correction lead to uncompensated decoherence in the tens to hundreds of MHz to GHz range, degrading overall coherence.

\begin{figure}[H] \label{fig:AS}
    \centering
    \makebox[\textwidth][c]{% ensures centering even if image is wider
    \includegraphics[width=1.0\textwidth]{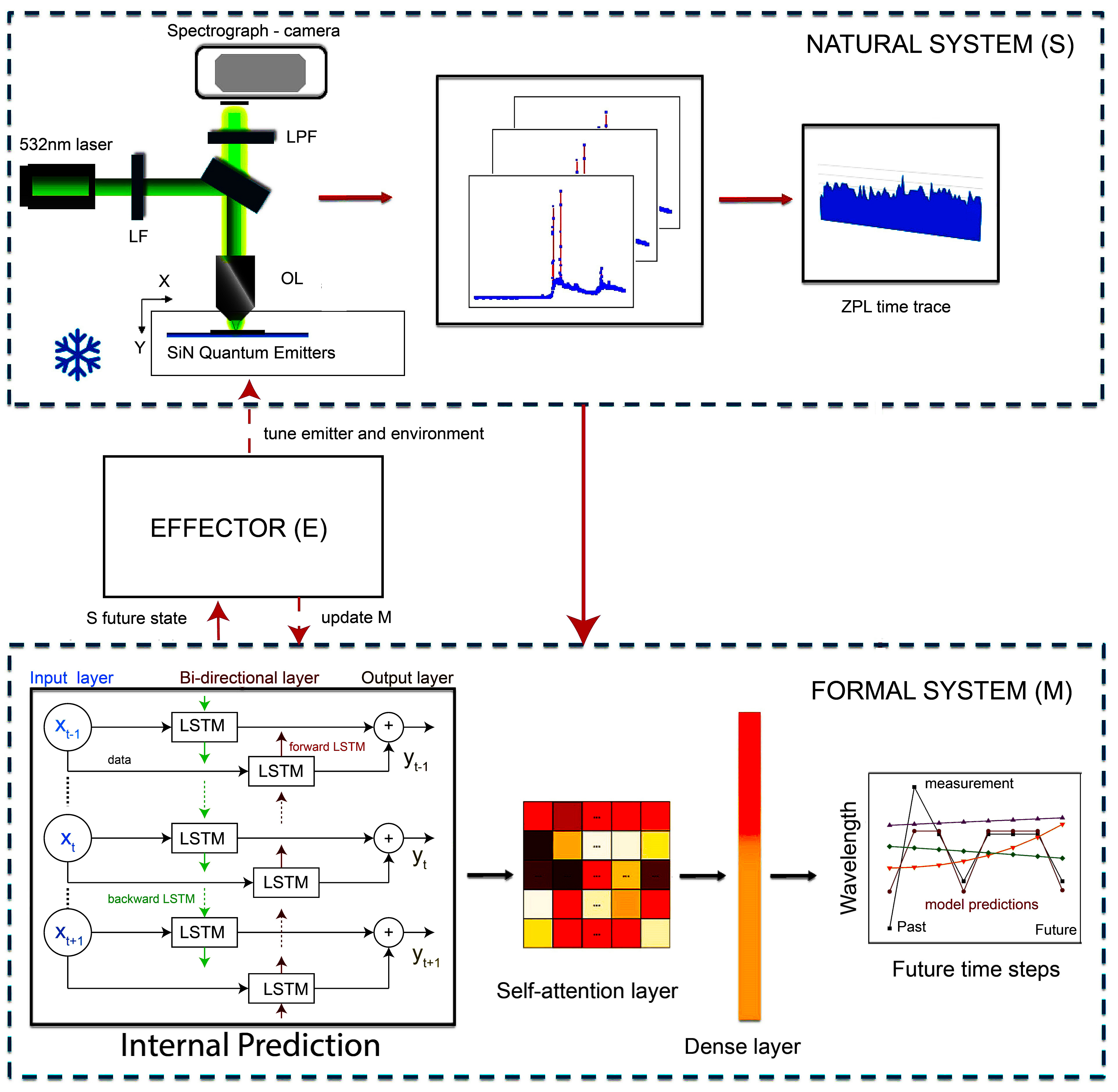}}
     \caption{$|$\textbf{Rosen/Nadin inspired Anticipatory Systems framework}: The natural system ($S$), consisting of a quantum emitter and environment (setup: OL: Objective Lens, LF: Laser cleaning filter, LPF: Long Pass Filter), modeled by a formal system ($M$) using an attention-based Bi-LSTM as its internal predictive core. $M$ updates the effector ($E$), which influences both $S$ and its local environment (the control line and effector feedback, marked with dashed maroon lines, will be implemented in our follow-up). The effector ($E$) also supplies contextual inputs, such as charge or phonon background (unrelated to $S$'s current state), to $M$. The internal model receives sequential spectral inputs $(x_{t-1}, x_t, x_{t+1})$, processed by a bidirectional LSTM—one branch capturing forward, the other backward temporal dependencies. Outputs $(y_{t-1}, y_t, y_{t+1})$ from both are concatenated and passed to a self-attention layer. A final dense layer integrates this for high-fidelity estimation of future states. The framework is platform-agnostic and could be applied to both optical and spin qubits. }\label{fig:experiment}
   \end{figure}

\noindent In this work, we move beyond reactive feedback-based schemes and show that decoherence in quantum systems with a non-Markovian signature exhibits temporally structured behavior that can, in fact, be anticipated. As a representative demonstration case, we investigate spectral diffusion in recently discovered intrinsic quantum emitters on the silicon nitride (SiN) platform~\cite{Martin2023}. At 4\,K, phonon contributions are strongly suppressed, rendering spectral diffusion one of the dominant manifestations of decoherence in our system. Therefore, we focus on spectral diffusion; the same statistical thermodynamic framework can equally be applied to other channels, such as spin dephasing in solid-state spin qubits and flux noise in superconducting circuits, where $1/f$-like noise and non-Markovian dynamics are likewise well established \cite{Oliver2023}. To understand the physical origin of this predictability, we employ Replica Theory (RP) and uncover temporal overlap structures in the emitter spectral trajectories, revealing intrinsic correlations that arise from slowly evolving environmental coupling~\cite{Parisi1983}. To further quantify these correlations and validate their persistence across various timescales, we perform autocorrelation and power spectral density analyses, observing $1/f$-type noise that confirms the presence of long-range memory effects and non-Markovian dynamics. Although prior efforts have attempted to predict the evolution of quantum system trends in narrow experimental regimes, and demonstrated predictive control in single-ion systems under artificially engineered noise \cite{Ramezani2024, Mavadia2017}, these approaches lacked universality across realistic open quantum systems. Our work bridges this gap by experimentally demonstrating that decoherence, here manifested as spectral diffusion, is not always purely stochastic and exhibits predictable features. Hence, its behavior can be learned across multiple distinct quantum systems and, ultimately, controlled.
\noindent We further adopt a framework inspired by anticipatory systems in biological networks~\cite{Rosen1985, Nadin01012010} and implement internal prediction through an attention-based bidirectional long short-term memory (Bi-LSTM) network~\cite{lecun2015deep, jing2020self} as shown in Fig.~\ref{fig:AS}. It is important to emphasize that, although the proposed architecture incorporates a control line and an effector, its operational principle is fundamentally distinct from conventional reactive feedback ~\cite{Sohn2019, acosta2012, Hansen2010, Kuruma2025}. Standard feedback schemes respond to measured deviations of the system state \emph{after} decoherence has already occurred, and are therefore inherently constrained by measurement back-action, feedback latency, and actuator bandwidth. By contrast, the anticipatory framework closes the control loop on inferred future system evolution and latent environmental dynamics, rather than on observed errors. This enables preemptive intervention prior to the onset of decoherence, substantially reducing sensitivity to latency and eliminating the need for continuous high-bandwidth measurement. In this regime, the effector operates on contextual environmental variables that determine subsequent system dynamics, rather than compensating for deviations in the measured system state. The proposed approach, therefore, falls outside the category of reactive feedback control.

\noindent An anticipatory system equipped with a trained Machine Learning (ML)-based internal predictive framework can accurately forecast spectral shift trends. This model learns bidirectional temporal correlations in the electronic transition of the quantum emitter (Zero Phonon Line-ZPL) evolution and generates high-fidelity forward predictions. The model captures emitter-specific memory effects while maintaining generalization across various quantum systems (For additional emitter data, refer to S3.3 \cite{pmaan2025}). Realization of the proposed predictive model could enable substantial suppression of spectral shifts, by factors ranging from approximately average value of 2.32 to 20.19 compared to systems without prediction, depending on the intrinsic stability of each emitter (see S4 \cite{pmaan2025}). Here, the reported improvement factor does not reflect anticipation accuracy. Instead, it quantifies the prospective reduction of step-wise ZPL excursions that would be achieved if the proposed predictive framework were implemented, relative to an
uncompensated reference. This metric, therefore, captures the potential for operational frequency stabilization enabled by the anticipatory framework.

\noindent Notably, the framework also offers a practical pathway for identifying and classifying quantum systems based on their temporal predictability, enabling exclusion of erratic or non-predictive emitters from the pipeline. This dual capability, encompassing both correction and pre-selection, offers a promising route toward the scalable deployment of solid-state quantum systems. To the best of our knowledge, this work presents the first experimental realization
of a replica-inspired, trajectory-resolved analysis for anticipatory inference in
non-equilibrium quantum dynamics. By experimentally validating a predictive
framework for spectral diffusion across multiple emitters, it establishes a
general foundation for coherence preservation and anticipatory control of
dynamical evolution in quantum systems.

\section{Results}
\subsection*{Spectral diffusion as a representative decoherence channel}

\begin{figure}[H]
    \centering
    \makebox[\textwidth][c]{% ensures centering even if image is wider
        \includegraphics[width=0.7\textwidth]{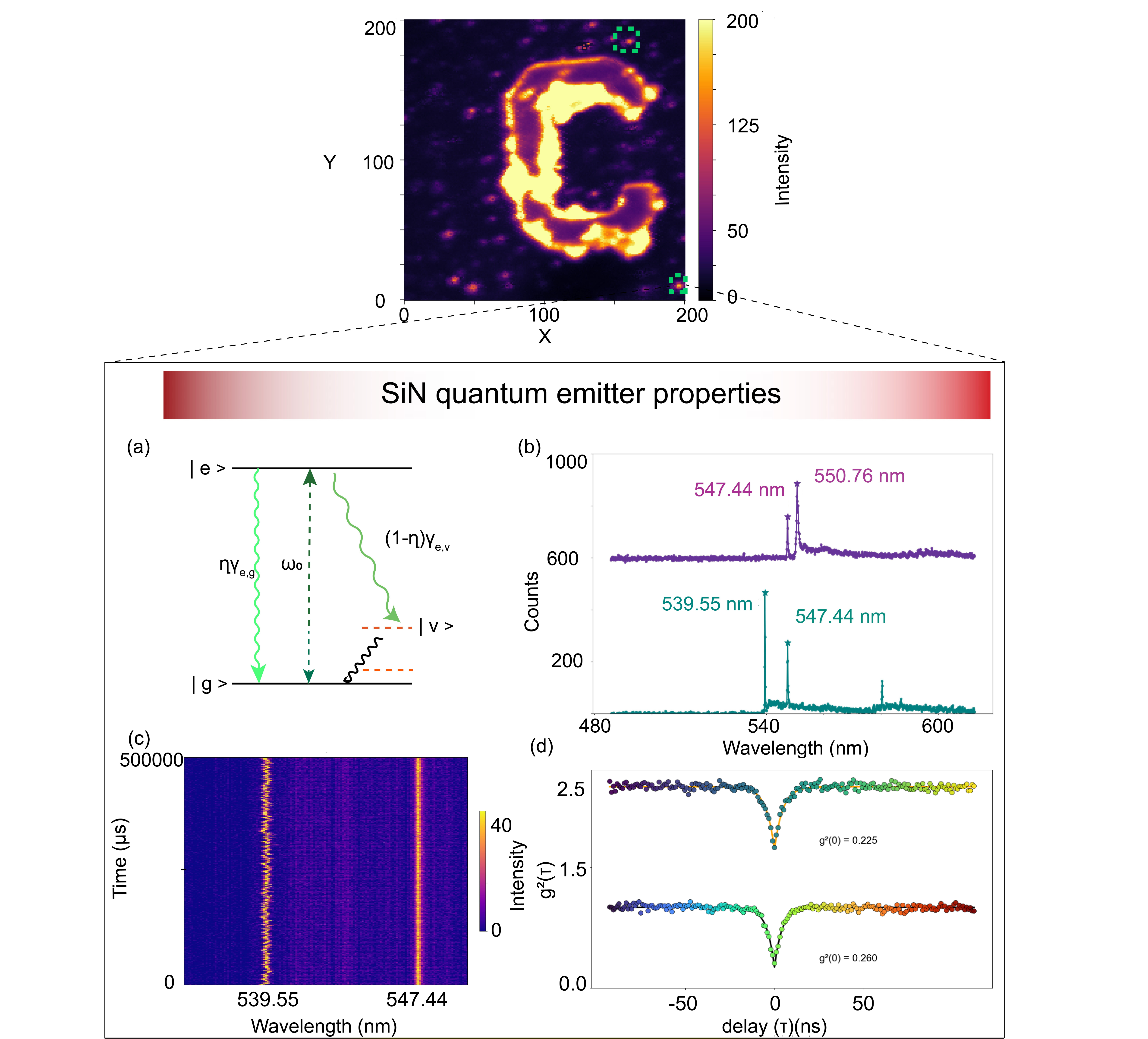}}
        \caption{%
    $|$ \textbf{Experimental characterization of decoherence dynamics in a quantum system}: 
    Non-resonant PL map (532nm excitation laser), showing two SiN quantum emitters within the dotted green circle:  
    (a) Energy level structure: Non-resonant PL (532nm excitation laser), the excitation laser frequency $\omega/2\pi$ is detuned from the transition frequency $\omega_0$ governed by the system Hamiltonian \( H_A \) with environmental-induced disorder.
 In contrast, during resonant excitation, $\omega_0$ matches one of the system eigenfrequencies.
 The radiative decay rate between states \( e \) and \( g \) is denoted by \(\eta \gamma_{e,g} \), for the excited to electronic and vibrational ground state, where $\eta$ is the Debye-Waller Franck-Condon factor, $v$ is the vibronic excited state,
    (b) Non-resonant PL spectra of two distinct SiN emitters, recorded with a $600\text{grooves}/\text{mm}$ grating and 5ms exposure: the Si Raman line at 547.44nm and zero-phonon lines (ZPLs) at 550.76nm and 539.55nm. A 532 nm dichroic mirror, in combination with a 532nm long-pass filter, was used to suppress the excitation laser. Plots have been shifted along the Y axis for better visibility.(c) PL spectra ($500\mathrm{\upmu}$s exposure) time trace illustrating the spectral diffusion of the emitter ZPL, at 539.55nm, alongside the relatively stable evolution of substrate-related Si-Raman peak at 547.44nm. This Si-Raman peak acts as a local sensor for experimental artifacts.  
    (d) Second-order autocorrelation functions $g^{(2)}(\tau)$ for two non-resonantly excited emitters that confirm their single-photon emission characteristics; all taken at $\approx 1$mW excitation power. Plots have been shifted along the y-axis for better visibility.}   
    \label{fig:PL}
\end{figure}

As an illustrative case, we examine spectral diffusion, using intrinsic quantum emitters on the silicon nitride (SiN) platform as the physical realization. Despite differences in how quantum emitters are formed, disorder remains an intrinsic consequence of the local environment. These effects cause time-dependent shifts in the emitter’s transition energy, limiting coherence and spectral stability~\cite{Seo2016, Hansen2010, spokoyny2020, Kuruma2025, santori2004}. In PL spectroscopy, carriers are injected above the bandgap and subsequently relax through myriad intermediate states before populating the zero‐phonon line, giving rise to a profusion of free charge carriers in the surrounding matrix. These carriers generate fluctuating local electric fields, in turn broaden and shift the emitter’s ZPL in PL spectra. In addition, the elevated pump power for non‐resonant excitation exacerbates local heating, further aggravating spectral diffusion. As shown in Fig.~\ref{fig:PL}, we use $\approx 1$mW non-resonant excitation wavelength at $532$nm to enhance the spectral wandering.
After deliberately maximizing environmental perturbations, we can characterize the temporal evolution of the ZPL centered around $\lambda_0^i(t) = (550.76\text{nm}, 539.55\text{nm})$ for two distinct emitters $i=1,2$, (shown in Fig. ~\ref{fig:PL}b), separated by $\approx20\upmu$m.
The temporal evolution of the ZPL was probed by collecting PL emission spectra under continuous-wave excitation (for additional emitter details, refer to S3.3 \cite{pmaan2025}). As depicted in Fig.~\ref{fig:PL} c, although the ZPL of the SiN emitter undergoes a pronounced spectral shift, the Si-Raman mode from the silicon substrate remains invariant, serving as an intrinsic reference that discriminates spurious spectral shifts due to experimental perturbations, such as exciting laser fluctuations, thermal drifts, and mechanical vibrations. Consequently, the invariance of the Si–Raman line confirms that the observed ZPL wandering is due to intrinsic emitter dynamics rather than systematic experimental changes (see~\ref{sec:spectral-diffusion}).

\clearpage
\subsection*{Replica theory formalism for decoherence }

We employ the RP to analyze one of the dominant decoherence channels, here spectral instability, defined through a decision variable that represents the deviation of the ZPL from its expected nominal value, \( \lambda_0^i(t)-\lambda_e \). This instability arises from disorder-induced spectral diffusion that can be modeled by a stochastic parameter \( D \), which captures the effect of local environmental fluctuations on the emission wavelength \( \lambda_0(t) \). Here, we consider a quantum emitter embedded in silicon nitride~\cite{Martin2023}, modeled as a three-level system as shown in Fig.~\ref{fig:PL}a, while the system-environment interaction is described by the Hamiltonian \( V_{\text{int}} \), scaled by a coupling parameter \( \beta \). 
Defining $A_i$ and $B_i^{(j)}$ as operators, acting respectively on the emitter and environmental degrees of freedom, the interaction Hamiltonian is expressed as $V_{\text{int}}=\sum_i\sum_j \beta_{i,j} A_i \otimes B_i^{(j)}$, where $j$ labels distinct decoherence channels of the $i^{\mathrm{th}}$ emitter. 

\noindent In open quantum systems, phenomenological models for memory effects and structured environmental interactions are analyzed under limited approximation using an effective memory-kernel non-Markovian master equation inspired by the Nakajima–Zwanzig formalism in the $\mathrm{Schr\ddot{o}dinger}$ picture~\cite[p.~78, ~82]{Rivas2011}
\begin{equation}
\begin{aligned}
i \frac{\text{d}\rho(t)}{\text{d}t} = 
\begin{aligned} 
& [H_A + \beta V_{\text{int}}, \rho(t)] +\zeta^2 \sum_j [H_{R}^j, \rho(t)]\\
&+i\zeta^2 \sum_\omega \sum_{m,n} \gamma_{mn}(\omega)\int_0^t\text{d}t' \kappa'(t,t')  \left[
\begin{aligned}
&A_n(\omega) \rho(t') A_m^\dagger(\omega)\\
&\quad-\frac{1}{2} \{A_m^\dagger(\omega) A_n(\omega), \rho(t')\} 
\end{aligned}\right]\end{aligned}\, ,
\end{aligned}
\end{equation}
where the kernel $\kappa'(t,t')$ characterizes the non-Markovian dynamics inherent in the phenomenological description of spectral diffusion~\cite{Rivas2011, Vacchini2024Open}. $H_A$ denotes the atomic Hamiltonian, $\rho(t)$ is the density matrix describing the quantum state of the emitter, and $H_R$ is the renormalization Hamiltonian capturing second-order perturbation due to interactions with the surrounding environment. This includes phonon fields or material defects that modulate the emitter’s transition energy over time, which is expressed as 
\( H[\lambda_0;D] = H_A[\lambda_0] + \beta V_{\text{int}}(D) + \zeta^2 H_{R}^j(D) \). he dissipative term captures the environment-induced decoherence governed by bath spectral density $\gamma_{m,n}(\omega)$.

\noindent  To capture the statistical properties of the solid-state system under disorder, we construct a partition function $Z$ that incorporates the effects of multiple decoherence channels modeled as a Gaussian process. This forms the basis for the replica approach, where different spectral trajectories of the emitter are treated as replicas of the same disordered system, all governed by identical disorder realizations \cite{Replica2021, FEYNMAN1963, Smith1987, MacKenzie2000} $Z=\int D[\lambda_0] \mathrm{e}^{-\beta H_A}\langle \mathrm{e}^{-\beta V_{\text{int}}(\text{bath},\lambda_0)}\rangle_{\text{bath}}$,
which, in the case of quenched disorder, leads to replica coupling due to averaging over a fixed but random environment.  These inter-replica correlations capture temporally structured environmental memory and appear as non-local interaction terms mediated by the memory kernel \( \kappa(t, t') \)
, and can be rewritten as $Z=\int D[\lambda_0] \mathrm{e}^{-\beta H_A} \mathrm{e}^{\frac{\beta^2}{2}\iint \mathrm{d}t\, \mathrm{d}t' \Delta\lambda_0(t)\Delta\lambda_0(t')\kappa(t,t')}$.
The disorder-associated free energy is computed using the replica trick, which is mathematically equivalent to averaging over multiple replicas of the emitter’s spectral trajectory, each containing a ZPL, subjected to the same disorder \( D \): \( \langle \ln Z[D] \rangle \).

\begin{figure}[H]
    \centering    
    \includegraphics[width=1.0\linewidth]{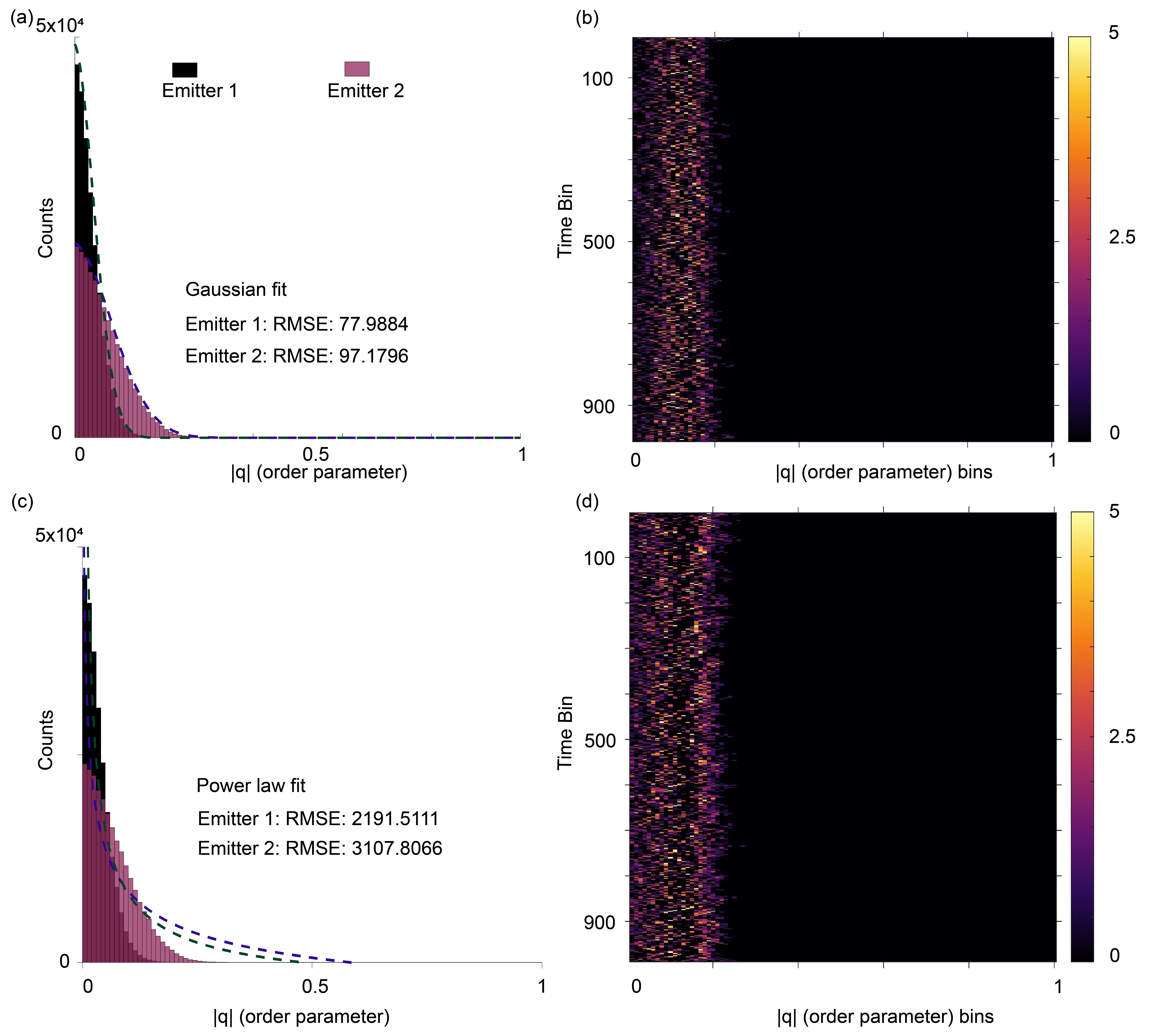}
    \caption{$|$ \textbf{Experimental results for decoherence channel analysis via an RP-based order parameter}: (a),(c) Spectral overlap represented by RP-based order parameter $|q|$ histogram for two different quantum emitters spectra. The overlap parameter, concentrated around 0, indicates low overlap for the spectral statistics. A non-zero value of the order parameter indicates a non-trivial overall overlap of the spectra. (b),(d) shows order parameter histogram evolution for two different quantum emitters under similar excitation conditions. The evolution of the overlap parameter across successive time bins, each consisting of 100 frames, with a sliding window shifted by 10 frames in the next row, reveals a dynamic temporal structure. This indicates that the overlap parameter is not a static quantity, but rather encodes time-dependent behavior. In an ideal scenario where all samples are identical, the overlap parameter would approach zero. However, in the present case, deviations from the average reference behavior yield non-zero overlap values, reflecting correlated fluctuations across different samples relative to the ensemble mean. Notably, relative stability is assessed via the stationarity of the fitted statistics for each time step. We define instability operationally as temporal reconfiguration of the statistical properties $\mu(t)$ or $\sigma(t)$ beyond their typical fluctuation bounds, resulting in diminished overlap between consecutive distributions.}
 (see S5 \cite{pmaan2025}) . While demonstrated here for spectral diffusion, the same predictive principle could be applicable to spin coherence in solid-state qubits and flux noise in superconducting circuits (see S6 \cite{pmaan2025}).
 \label{fig:order}
\end{figure}
\noindent To quantify the similarity between full spectral trajectories over time, including the ZPL, we define an overlap matrix. Under the assumption of the replica symmetry, this matrix satisfies \( q_{aa}(t,t') = 1 \), indicating perfect self-similarity, and \( q_{ab}(t,t') = q_{ba}(t,t') \) for \( a \neq b \), reflecting that replicas are statistically equivalent, with the following explicit form 
\begin{equation}\label{eq:overlapMatrix}
q_{ab}(t, t') = \frac{
\langle \Delta_a , \Delta_b \rangle
}{
\left\| \Delta_a \right\| \, \left\| \Delta_b \right\| }\, 
\end{equation}
where the mean spectral trajectory across replicas is defined as \( \bar{\lambda}_0(t) = \frac{1}{n} \sum_{i=1}^n \lambda_0^{(i)}(t) \), and serves as the reference for evaluating the deviation of the individual frames, $\Delta_i=\lambda_0^{(i)}(t) - \bar{\lambda}_0$, $\left( i=\{a,b\}\right)$, and then correlations between all possible frame pairs in the overlap (order parameter) matrix \eqref{eq:overlapMatrix}. We compute the histogram of the order parameter matrix for two distinct quantum emitters (see S2), as shown in Fig.~\ref{fig:order}a, to quantify the distribution of spectral overlap values across replicas over time. 
This analysis reveals the degree of temporal correlation between disorder-induced spectral trajectories. The presence of non-zero \( q \) values (up to \( \sim0.2 \)) indicates that the replicas are not entirely uncorrelated, suggesting an underlying temporal structure in the spectral diffusion dynamics. The replica framework in Fig.~\ref{fig:order}a formally assumes quenched disorder and an equilibrium description, whereas spectral diffusion in solid-state emitters is intrinsically non-equilibrium and evolves on experimental timescales on the scale ranging from $\upmu$s to s. For this reason, we primarily analyze the time evolution of the order parameter to capture departures from stationarity. When the order parameter is instead evaluated over the entire
dataset, the analysis effectively coarse-grains across many quasi-quenched disorder realizations (see S1 \cite{pmaan2025} for definition) and integrates over slow environmental fluctuations. At this level of
coarse-graining, the resulting distribution is well described by a Gaussian, consistent with the central-limit behavior expected when many finite-variance contributions are aggregated. The significantly poorer power-law fits indicate the absence of heavy-tailed statistics at this global scale, although they do not preclude non-equilibrium or intermittent dynamics at shorter timescales, which are resolved in Fig.~\ref{fig:order}b.

\noindent To perform a trajectory-resolved analysis of non-equilibrium quantum dynamics, we
exploit a quasi-quenched regime in which fast emitter fluctuations are conditioned
on slowly evolving disorder configurations. By experimentally engineering this
timescale separation, we demonstrate a trajectory-resolved framework that reveals
dynamical structure inaccessible to conventional correlation-based or purely
spectral methods (see S1\cite{pmaan2025}). We further define a temporal window (see Fig. S5 \cite{pmaan2025} with Frame size = 10 samples) within which the slow disorder can be treated as effectively quenched. This framework enables us to elucidate the temporal evolution of inter-replica correlations with greater precision. To quantitatively assess these correlations, we compute the order parameter $|q|$ for each time frame and construct histograms of $|q|$ values (see S4 \cite{pmaan2025}). As shown in Fig.~\ref{fig:order}b and c, fluctuations in the order parameter reflect an underlying temporal structure in the spectral dynamics, indicating non-trivial correlations in the system’s response to disorder. The intermittent excursions of the order parameter toward higher values could reflect increased step-to-step variability in the underlying dynamics, which in turn could limit predictability of the subsequent evolution. Therefore, in our empirical observations, emitters exhibiting larger fluctuations in the time evolution of the order parameter were associated with reduced predictability, whereas emitters whose order parameter remained more tightly bounded about its mean exhibited comparatively higher predictability. In such regimes, the conditional distribution of future states broadens, reducing the extent to which anticipatory prediction can prospectively constrain spectral drift (see S5\cite{pmaan2025}).

\textbf{Decoherence channel noise analysis}\label{sec:spectral-diffusion}

Although decoherence pathways can be phenomenologically explained~\cite{Rivas2011}, the exact source of memory effects in solid-state platforms is still speculative. For example, in quantum dots, memory effect arise from defects that randomly hop between two metastable states. These defects are often modeled as two‐state fluctuators, each characterized by distinct transition rates~\cite{paladino2014}. When such a group of fluctuators with a broad rate distribution acts collectively, environmental noise (e.g., charge fluctuations) induces shifts in the ZPL of the emitter.  These shifts manifest themselves as spectral fluctuations in the emitted photon’s wave packet with $1/f$ signature and electric‐field envelope~\cite{paladino2014}:
$E(t) \propto A(t)\,\mathrm{e}^{\,i\bigl[\omega_0 t + \varphi(t)\bigr]},
\qquad \varphi(t) = \int_0^t \Delta\omega(t')\,\mathrm{d}t'$, where $\omega_0$ is the central ZPL frequency and $\Delta\omega(t)$ is the instantaneous frequency shift, i.e., the spectral diffusion. Another intuitive picture for the origin of $1/f^\alpha$ noise in these systems employs simple two‐ or three‐level models.  In an ideal two‐level emitter, the spectrum is purely Lorentzian.  If the excited‐state energy undergoes slow fluctuations due to, e.g., local electromagnetic or vibrational noise over an observation period $t$, the instantaneous spectrum would be a weighted superposition of Lorentzians with varying center frequencies and linewidths.  If the fluctuation timescales $T$ are continuously distributed and $T \leq t$, their cumulative effect naturally yields a $1/f^\alpha$ Power Spectral Density (PSD) when averaged over many realizations~\cite{Kaulakys2005}.

\begin{figure}[H]
    \centering
    \includegraphics[width=1.0\linewidth]{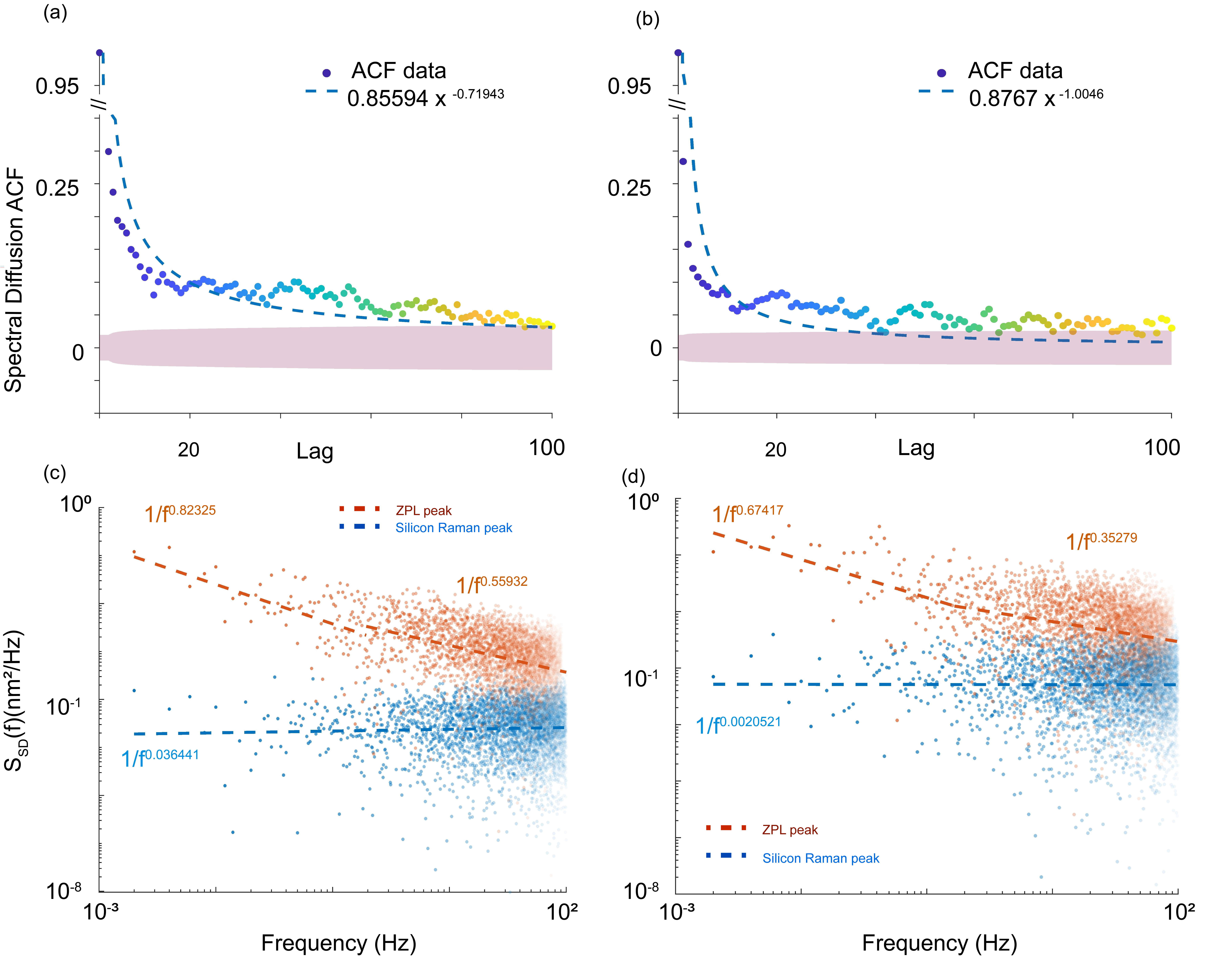}
    \caption{$|$ \textbf{Decoherence channel noise analysis}: (a),(b) Autocorrelation function (ACF) of the ZPL time traces for two quantum emitters, where each lag represents a temporal shift relative to itself. Slow decay in the ACF indicates non-trivial temporal correlations, while fluctuations beyond lag $>40$ for emitter (b) alternate between significant and noise-like. The shaded region shows the 95\% confidence interval under the null hypothesis of no correlation; values outside this region indicate meaningful correlations. Statistical bounds were calculated as $r(k) \pm 1.96 \sqrt{\frac{1}{N} \left(1 + 2 \sum_{i=1}^{k-1} r(i)^2 \right)}$,
    where $N$ is the sample size. (c),(d) Power spectral density (PSD) of the ZPL traces over 10,000 spectra (individual exposure time} $500\,\upmu$s). The absence of $1/f$ behavior in the Si-Raman line confirms the noise is not instrumental, while the presence of $1/f$ components in the ZPL indicates long-range temporal correlations consistent with non-Markovian emitter–environment interactions. At 4\,K, phonon contributions are suppressed, making spectral diffusion a dominant decoherence channel in this demonstration; the framework is general and applicable to other noise processes.
    \label{fig:noise_final1}
\end{figure}

\noindent To isolate noise contributions arising from the experimental setup, including laser and optical instabilities, mechanical vibrations, detector noise, and temperature fluctuations, we simultaneously monitored the Silicon Raman line (a built-in feature of the substrate) alongside the ZPL emission (Fig.~\ref{fig:noise_final1})c,d and computed the corresponding PSD. Fitting the low-frequency regime of the PSD for the ZPL revealed a clear $1/f^\alpha$ dependence with $\alpha > 0.5$, whereas the high-frequency slope reduced for both the quantum emitters to $0.55932$ and $0.35279$ respectively, indicating reduced temporal correlations. In all the emitters, the Si-Raman line consistently showed $\alpha \approx 0$. The absence of a $1/f$ signature in the Si-Raman line, which serves as a local probe, confirms that the observed memory effects are intrinsic to the quantum emitter.
Finally, to confirm the presence of memory effects in spectral diffusion, we calculate the autocorrelation function (ACF) of the ZPL wavelength time series Fig.~\ref{fig:noise_final1}a,b. The ACF quantifies how strongly the spectral trajectory of the ZPL at a given time point correlates with itself at later times, defined by the time lag. For each lag value, the ZPL time series is temporally shifted and compared with its original version; sustained correlations at non-zero lags indicate that past spectral values influence future ones-an unambiguous signature of memory. Figure~\ref{fig:noise_final1}a, b shows the ACFs for two emitters, evaluated over 100 lags. The persistence of nonzero ACF values with a power-law fit of -0.71943, indicating stronger temporal correlation, and -1.0046, exhibiting relatively weaker correlation, confirms the presence of long-term temporal correlations-a hallmark of non-Markovian dynamics. 

\subsection*{Anticipatory system representation for predicting decoherence}

Dynamic decoupling (DD) is highly effective for suppressing stationary, weakly correlated noise, particularly when the environment can be well approximated as Gaussian and Markovian~\cite{Lloyd1998, Lidar2005, Choi2020, Suter2011}. However, in DD protocols, the effective Hamiltonian is generated via the Magnus expansion, e.g.\ $H_{\mathrm{eff}}^{(1)} \sim \frac{1}{T}\int_0^T\!\mathrm{d}t_1\int_0^{t_1}\!\mathrm{d}t_2\,[\tilde H(t_1),\tilde H(t_2)]$, such that their higher-order terms involve nested commutators, creating an expanding set of system-bath operators that grow rapidly with interaction order. As a result, the practical complexity of DD increases sharply (see S2~\cite{pmaan2025}).

\noindent Optimal control methods assume a fixed Hamiltonian during optimization\cite{Brif2010}.  In the presence of persistent parameter drift, the tracking error between the applied control and the instantaneous optimum obeys a recursion of the form $\|e_k\|\le \rho^k\|e_0\|+\Delta \frac{1-\rho^k}{1-\rho}$ and therefore does not vanish asymptotically, but instead approaches a finite bound set by the drift magnitude (see S2~\cite{pmaan2025}). Consequently, offline open-loop optimal control cannot asymptotically track the true system optimum under non-stationary conditions, placing a fundamental limit on the achievable control fidelity.

\noindent We therefore employed a replica-theory-based analysis, which circumvents the need to explicitly model microscopic bath dynamics, to track the temporal evolution of the overlap between configurations. The emergence of nontrivial overlap dynamics demonstrates that spectral diffusion is temporally structured rather than random, providing a physical basis for the use of an anticipatory architecture. Anticipatory systems, originally formulated for biological contexts \cite{Rosen1985, Nadin01012010}, provide a powerful framework for designing edge-inference hardware capable of long-range prediction and active control of spectral diffusion in quantum emitters. Figure~\ref{fig:AS} illustrates our architecture, where the natural system $S$, comprising a single-photon emitter and its local environment, co-evolves with its formal model $M$ over a time interval $T$, enabling synchronized inference and control. By tracking $S$ in real time, $M$ effectively projects the system's future state and generates predictions that could drive the effector $E$, implemented on a local processor, to apply preemptive stimuli to $S$. In the envisioned hardware implementation, $E$ will both enact these stimuli on the emitter and environment and update $M$ to incorporate the resulting system changes.

\noindent The efficacy of this anticipatory framework depends critically on the fidelity of the encoding map between \(S\) and \(M\). To ensure that the evolutions of \(S\) and \(M\) commute, we propose continuous recalibration of the encoding via (1) data‑driven reconstruction of historical and predicted mappings to maintain the internal coherence; (2) adaptive learning that incrementally refines the model in real time as context shifts; and (3) feature engineering aimed at expanding representational capacity for richer environmental sensitivity. We refer to this as Rosen-Nadin-Inspired Continuous Recalibration (R-NICR), meaning the continuous re-identification of the system--model encoding map under slowly drifting environment. This procedure does not constitute a feedback that corrects observed system errors, but instead maintains causal consistency between the internal model’s inferred context and the system’s subsequent evolution, enabling meaningful comparison between the trajectories of $S$ and $M$ over time~\cite{Rosen1985, Nadin01012010}

\subsection*{Predicting decoherence in quantum systems}

The developed internal predictive model, within the anticipatory framework, draws inspiration from the Atkinson–Shiffrin memory model~\cite{ATKINSON1968}, which is implemented through a Recurrent Neural Network (RNN) framework utilizing Long Short-Term Memory (LSTM) units. In particular, the bidirectional LSTM architecture is naturally suited to model such forward–backward temporal dependencies and embedded memory. 
 
\begin{figure}[H]
    \centering
    \includegraphics[width=1.04\linewidth]{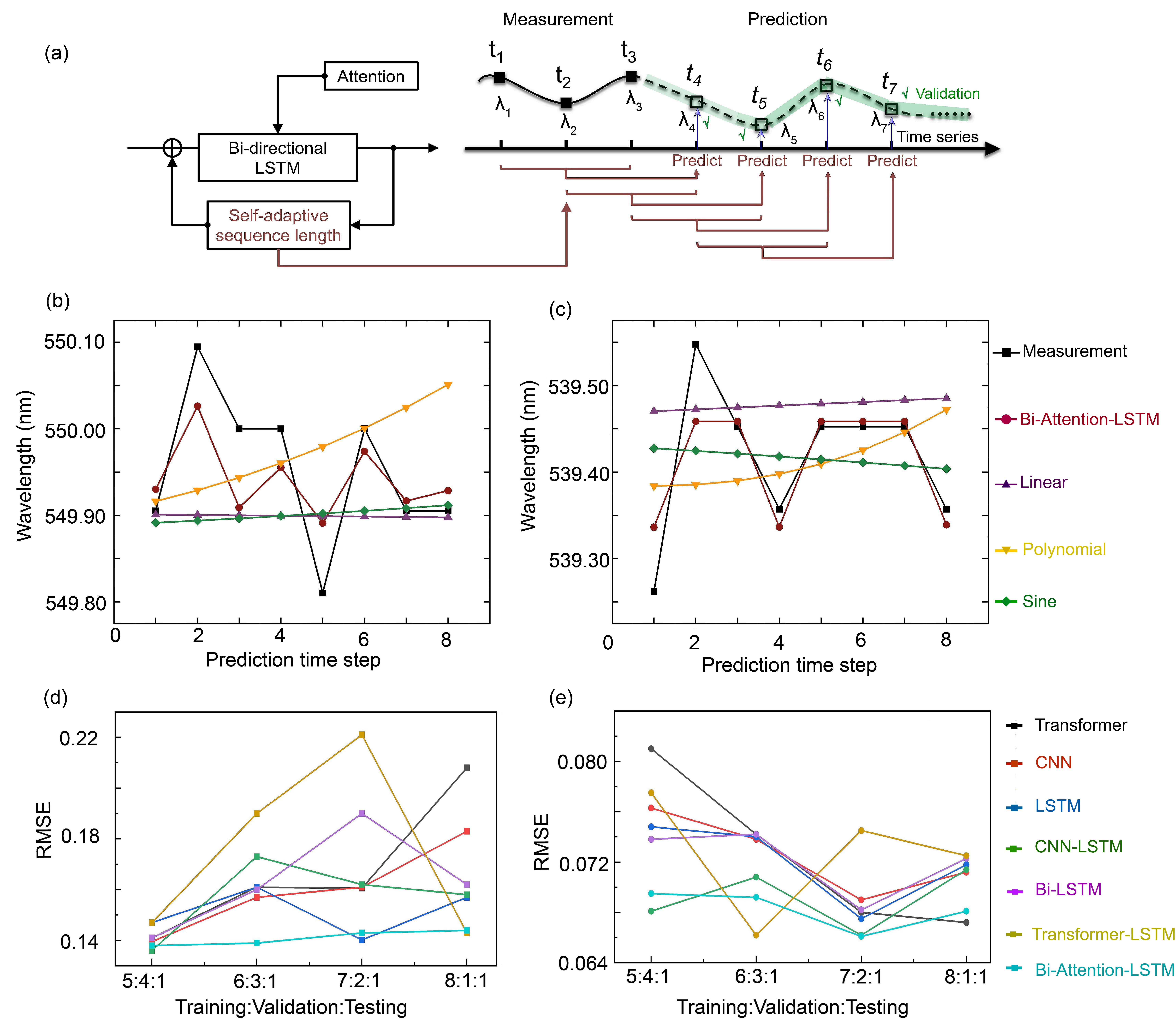}
    \caption{$|$ \textbf{Anticipatory prediction architecture and decoherence prediction:} (a) The model receives a time series of ZPL measurements up to time \( t_3 \). The self-adaptive sequence length mechanism dynamically determines the effective input window at each step. An attention mechanism then assigns dynamic weights to different time steps. Predicted values \( \lambda_p(t_k) \) are recursively fed back as inputs for subsequent time steps, forming an auto-regressive prediction loop that internalizes forward temporal dynamics. The green region represents the tolerance for a valid prediction. (b),(c) Spectral wavelength diffusion predictions comparison for two different quantum emitters, with data acquired at 500 $\mathrm{\upmu}$s intervals. The anticipatory model (Bi-Attention-LSTM) closely tracks the measured fluctuations, while linear, polynomial, and sine-function predictors fail to capture the underlying nonlinear dynamics. (d),(e) Quantitative RMSE (Root Mean Square Error) loss performance comparison of time-series prediction models under different dataset partitioning schemes for two distinct emitters.}
    \label{fig:common_caption}
\end{figure}

\noindent We predict future ZPL values $\lambda_p(t_k)$ at a given time $t_k$ using an attention-enhanced bidirectional LSTM architecture, as outlined in~\ref{alg:Attention_BiLSTM}. This architecture minimizes the validation loss $L^*_{\text{val}}$ and generates an optimal model $\mathcal{M}^*$. Figure~\ref{fig:common_caption}(a) shows that historical measurements up to $t_3$ are given as input to the encoder module, initializing future predictions. Following each prediction, a newly inferred value $\lambda_p(t_k)$ is appended to the input sequence, enabling recursive and autoregressive prediction with no additional external measurements. A transformation, $x_{\text{input}} \rightarrow \text{Bi-LSTM Encoder} \rightarrow \mathbf{H}_e,(\mathbf{H}_e, \mathbf{h}_{\text{dec}}) \rightarrow \text{Attention Mechanism} \rightarrow \mathbf{C}_e,
\mathbf{C}_e \rightarrow \text{Fully Connected Layer} \rightarrow \hat{y}_e,$ governs this recursive forecasting loop, where $\mathbf{H}_e$ denotes the encoded hidden states, $\mathbf{h}_{\text{dec}}$ is the decoder’s internal state, and $\mathbf{C}_e$ is the context vector constructed by the attention weights.

\noindent The attention mechanism computes dynamic relevance scores for each time step in $\mathbf{H}_e$, producing attention weights $\alpha_i$. The context vector is then computed as ~$\mathbf{C}_e=\sum_{i=1}^{L} \alpha_i h_i$,
where $\alpha_i$ are dynamically computed attention weights and $h_i$ are hidden vectors in each time step $i$. This formulation enables the model to dynamically reweight historical latent representations at time \( t \) according to their contextual relevance, thereby capturing and internalizing the temporal dependencies underlying the predictive structure of the time series.

\noindent Importantly, the attention mechanism implicitly rescales each encoder state by modulating its contribution to the context vector $\mathbf{C}_e$, effectively contracting or expanding the memory horizon of the model. This capacity enables the model to adapt to long-range dependencies or abrupt regime changes. Simultaneously, a self-adaptive sequence length mechanism dynamically adjusts the effective temporal window—shortening or lengthening the input history based on current signal volatility—thereby enhancing robustness to evolving background fluctuations and mitigating the effects of non-stationary behavior. Through repeated autoregressive updates, the network constructs an internal forward model of the underlying physical dynamics. 

\noindent We successfully predicted up to 8 future spectral occurrences for two distinct quantum emitters, (see Fig.~\ref{fig:common_caption}(b),(c) and S3.3 for additional emitters). The black curves represent the experimentally measured spectral diffusion trajectories, while the remaining curves show predictions obtained using different forecasting models: Bi-Attention-LSTM, linear regression, $5^{th}$ order polynomial fitting, and sine function approximation. The Bi-Attention-LSTM model demonstrates superior predictive performance, accurately capturing both the amplitude and frequency of fluctuations present in the experimental data. This is achieved through its ability to model non-linear temporal dependencies and incorporate bidirectional context, which is critical for systems influenced by complex stochastic and memory-dependent processes. This forward-looking, data-driven behavior further reinforces the model’s anticipatory nature, distinguishing it from reactive models that rely solely on past data without internal simulation of future states.

\noindent  For the more stable emitter (see S4 \cite{pmaan2025}), as shown in Fig.~\ref{fig:common_caption}(c), the largest spectral mismatch in the experimentally measured trajectory occurs between the first and second time steps, with a spectral shift from 539.2619 nm to 539.5476 nm, corresponding to a mismatch of approximately $294$ GHz. In contrast, the Bi-Attention-LSTM model, leveraging its predictive capability, reduces the mismatch during the same interval to approximately $98$ GHz, achieving nearly a threefold improvement. In a representative case at time steps 4 and 5, the experimental mismatch is approximately $98.1$ GHz, whereas the model prediction reduces this to $6.18$ GHz, reflecting a 15.8-fold improvement. Similarly, for the relatively less stable emitter shown in Fig.~\ref{fig:common_caption}(b), the spectral mismatch in the experimental trajectory at time step 2, where the wavelength shifts from $550.0948$ nm to $550$ nm, corresponding to a mismatch of approximately $94$ GHz. When Bi-Attention LSTM-based prediction is applied, the predicted wavelength at the same time step improves to $549.9553 $ nm at step 5, reducing the spectral mismatch to $44.3$ GHz, an improvement by factor of approximately $2.1\times$. These results demonstrate the potential reduction in spectral mismatch achievable through the anticipatory framework, had it been implemented in real time.
 In contrast, traditional models fail to reproduce the dynamics observed in the measurements. The linear model consistently underfits the measured data, as a comparison, unable to account for sudden transitions. Polynomial fitting introduces unrealistic long-term trends and diverges beyond the initial prediction window. The sine function approximation imposes an artificial periodicity that does not reflect the underlying stochastic behavior of the system. To empirically validate this advantage, we benchmarked Bi‑Attention‑LSTM against a suite of open‑source time‑series predictors under four partition schemes (5:4:1, 6:3:1, 7:2:1 and 8:1:1). These ratios represent the relative proportions of training, validation, and testing data, respectively, allowing us to assess each model's performance and generalization ability under varying levels of data availability and distribution. Bi‑Attention‑LSTM demonstrates consistently low and stable RMSE (Root Mean Square Error) across all partition schemes for both emitter configurations, outperforming or nearly matching alternative architectures in nearly every case.

%% Transformer‑based and hybrid models (e.g., Transformer‑LSTM) achieve competitive minima but suffer from pronounced variability—most notably under the 7:2:1 split—undermining their reliability. Pure CNN, LSTM, and CNN‑LSTM designs exhibit moderate accuracy but fluctuate significantly with training‑set size, indicating sensitivity to data availability. In contrast, Bi‑Attention‑LSTM’s bidirectional recurrence coupled with attention‑enhanced temporal encoding confers robust generalization and minimized error across diverse data partitions, making it a premier candidate for forecasting emitter‑wavelength fluctuations. 

\noindent Transformer-based models and their hybrid counterparts, such as Transformer-LSTM, achieve competitive RMSE values, particularly under specific data splits like 5:4:1 and 6:3:1, as shown in Fig.~\ref{fig:common_caption}(d),(e). For example, the Transformer-LSTM reaches RMSE values as low as 0.148 at 5:4:1 and 0.066 at 6:3:1 in (d) and (e), respectively. However, their performance is notably inconsistent, with significant spikes under the 7:2:1 data split. In this case, the Transformer-LSTM's RMSE increases sharply to 0.221 (d) and 0.075 (e), the highest among all models depicted. Such variability suggests that although these models are capable of high prediction accuracy, they are particularly sensitive to changes in data partitioning, limiting their reliability in practice.

\noindent In contrast, traditional architectures like CNN, LSTM, and CNN-LSTM demonstrate moderate RMSE values—defined here as approximately 0.14–0.18 in Fig.~\ref{fig:common_caption}(d) and 0.068–0.075 in Fig.~\ref{fig:common_caption}(e). However, their accuracy fluctuates substantially across different train-validation-test splits. For instance, in Fig.~\ref{fig:common_caption}(d), the CNN model exhibits an RMSE rise from around 0.14 at the 5:4:1 split to 0.18 at 8:1:1, a variation of 28.57\%. This trend indicates that these models are less robust to shifts in training set size and may underperform when data availability is constrained or non-uniform.

\noindent Notably, the Bi-Attention-LSTM model, which combines bidirectional LSTM processing with attention-based temporal encoding, delivers the most stable and accurate performance across all examined data splits. In both subplots of Fig.~\ref{fig:common_caption}(d),(e), it consistently achieves one of the lowest RMSE, approximately 0.14 and 0.068, respectively, with minimal variation (less than 5\%) across the full range of splits. This consistent performance highlights its strong generalization capacity and positions it as the most robust and reliable model for forecasting emitter-wavelength fluctuations in diverse data scenarios.

\noindent Overall, we observed that the model effectively captured spectral shifts for both emitters, yielding a reduction by factors of up to $15.8\times$ for a relatively stable emitter and about $2.1\times$ for a much less stable one.  

\section{Discussion}
In this work, we comprehensively analyzed one of the dominant decoherence channels in a quantum system (spectral diffusion in multiple SiN quantum emitters as a representative use case here). Subsequently, we formalized this stochastic dynamics within a replica‐theory framework, thereby enabling a quantitative description of the emitter-to-emitter variability, specifically the temporal evolution of ZPL, correlation of the ZPL time trace, and relative fluctuation of the spectrum compared to its average behavior. We further demonstrated pronounced temporal correlations in the ZPL emission and identified a clear long-term memory effect, as evidenced by a robust power-law signature in the power spectral density. To address the challenge of predicting such complex, non-Markovian behavior across unseen, diverse emitters, we introduced a novel paradigm inspired by anticipatory systems and validated a formal system model driven by a machine-learning prediction algorithm. Experimentally, we exploited the intrinsic Si–Raman line from the silicon substrate as an in situ noise sensor, enabling us to isolate and quantify local contributions from laser intensity fluctuations, thermal instabilities, and micromechanical vibrations. Although reactive feedback–based control schemes have been shown to mitigate spectral diffusion effectively \cite{Sohn2019}, our anticipatory approach offers, in principle, the capacity to suppress any residual diffusion across emitters, thereby enhancing the coherence stability of quantum photonic devices.

\noindent The diverse set of quantum emitters characterized by ZPL wavelengths at 550.76 nm and 539.55 nm, focusing on two representative cases: one exhibiting typical temporal behavior and another showing pronounced instability, with a PSD approaching the white noise limit at higher frequency. Variability across emitters arises from differences in intrinsic ZPL wavelengths and local environments, both of which strongly influence the temporal dynamics of ZPL emission. These variations manifest in distinct temporal correlation features: one emitter exhibits an ACF with significant correlation extending up to approximately 85 lags, whereas the other shows correlation only up to around 40 lags. Differences are also evident in the evolution of the order parameter trajectories: the more stable emitter displays relatively infrequent but larger overlap changes, while the less stable emitter shows more frequent fluctuations with lower overlap, indicative of greater temporal variability. Long-range correlations further distinguish the emitters: while one exhibits a PSD profile consistent with $1/f$-like noise, indicating long-memory dynamics, the other approaches the white-noise limit, particularly at higher frequencies. Crucially, our framework accommodates this emitter-to-emitter heterogeneity and demonstrates robustness and scalability across a broad range of spectral and dynamical behaviors, enabling reliable characterization and predictive modeling of diverse quantum emitters. 

\noindent While our approach demonstrates robustness under worst-case spectral diffusion conditions, its applicability under resonant, low-power excitation, particularly in regimes approaching lifetime-limited linewidths but still influenced by residual spectral diffusion, remains to be experimentally explored to assess whether anticipation can be sustained. Nevertheless, the present results establish that this anticipatory framework can reliably enhance spectral stability. This demonstration paves the way for scalable quantum platforms by showing that our anticipatory approach not only could enhance relative spectral matching—and, consequently, mutual quantum coherence—across large arrays of quantum-emitter nodes, but also establishes a foundation for self-stabilizing, hybrid quantum–classical networks with optimized inter-node parameters.  More broadly, by explicitly addressing emitter heterogeneity, local environmental variability, and runtime fluctuations in multi-emitter systems, our methodology may enable a new class of fault-tolerant, self-synchronizing quantum–classical architectures in which precise qubit–bit synchronization across distributed nodes is essential~\cite{steinmetz2019}.  As part of future work, we aim to improve the prediction accuracy by investigating dominant decoherence mechanisms in greater depth. In parallel, we plan to enable Hamiltonian learning and achieve control of the local environment through electrical and strain tuning, while accounting for the time latency required to enable anticipatory control. Finally, we will realize the system on hardware with edge inference capabilities.~\cite{MAAN2022, PMAAN2022, 10.5555/3495724.3496706,10.1145/3007787.3001163, 10.5555/2969239.2969366, 10.5555/3495724.3497367}.

\noindent The anticipatory framework is designed to operate through continuous monitoring of environmental bath information, such as phonon or charge-bath dynamics, inferred from the acquired spectra and used as contextual input to an internal model. Based on this inferred context, the model generates calibrated actuation signals to perturb either the quantum system or its environment. The resulting system response is then re-measured, thereby closing the sensing–inference–actuation loop.

A central challenge in implementing this framework is the latency and bandwidth of the actuation pathway. The approach requires high timing accuracy and ultra-low-latency hardware pipelines capable of supporting real-time inference and control. To meet these requirements, the proposed architecture employs an FPGA-based effector, which enables ultra-low-latency, high-bandwidth signal processing and model-informed actuation. In this configuration, the FPGA performs real-time signal processing and inference to maintain temporal alignment between the internal model \(M\) and the physical system \(S\). In addition, propagation delays between sensing, model update, and actuation must be explicitly characterized and compensated to ensure temporal consistency across the full control loop.

\noindent Although we validate our framework in the context of optical decoherence via spectral diffusion as a representative decoherence channel, the replica-theory formalism and anticipatory architecture are platform-agnostic (see S6\cite{pmaan2025}). They could be directly applied to spin-based decoherence in solid-state systems such as hexagonal boron nitride (hBN), Silicon-Vacancy (SiV) centers in diamond, and quantum dots, where spectral variability and emitter-to-emitter heterogeneity remain key challenges. The same framework could also be used to characterize spin-bath dynamics in NV centers, quantum-dot qubits, and flux noise in superconducting circuits. This universality highlights that our contribution is not emitter-specific, but provides a generalizable route to predictive decoherence engineering.

\section{Methods}\label{sec2}
\subsection{Sample preparation}
The silicon nitride quantum emitter sample was prepared following a process reported in Ref.~\cite{Martin2023} and outlined in S7 \cite{pmaan2025}. A 100 nm layer of $\mathrm{SiO}_2$ was deposited using High-Density Plasma Chemical Vapor Deposition (HDPCVD, Plasma-Therm Apex SLR) technique to achieve higher plasma density than inductively coupled plasma source based Plasma-Enhanced Chemical Vapor Deposition (PECVD), followed by the deposition of a 200 nm SiN layer. The sample was then subjected to rapid thermal annealing (RTA) at $1100\,^\circ\mathrm{C}$ for 120\,s in a nitrogen atmosphere using a Jipelec JetFirst 200C RTA system. Finally, alignment markers were patterned using photolithography technique.

\subsection{Experimental setup}
Low-temperature PL characterization of the SPEs was performed using a confocal microscope integrated into a closed-cycle Montana S100 cryostation, stabilized at $\approx$ 4K  (Fig ~\ref{fig:experiment}). The sample  was non-resonantly excited using a continuous-wave Cobolt diode laser : Cobolt 08-DPL (filtered through a 532nm notch filter) through a 100$\times$ in-vacuum objective (Zeiss, NA = 0.85). The PL map was obtained by filtering the 532 nm excitation laser using a 532 nm dichroic mirror and a 532 nm longpass filter, using a two-axis galvo scanner integrated into a 4f imaging system. Quantum emitters are observed around the marker "C" (Fig.~\ref{fig:PL}). Photon anti-bunching was measured using a 50:50 fiber beam splitter and a pair of large-area superconducting nanowire single-photon detectors (SNSPDs), connected to a PicoQuant HydraHarp time-correlated single-photon counting (TCSPC) module for time-tagging photon events with acquisition resolution of 1.024ns.  A Princeton Instruments IsoPlane SCT-320 spectrograph, coupled with a PIXIS 400BR eXcelon camera, was used for PL spectral measurements with a 600 grooves/mm grating.

\subsection{Statistical information}
The $g^{(2)}$ fit was obtained using the following expression:
\[
g^{(2)}(t) = g^{(2)}_{\infty} \left(1 - \left(1 - \frac{g^{(2)}_0}{g^{(2)}_{\infty}} \right) \mathrm{e}^{-t_{\text{adj}} / \tau_{\text{antibunch}}} + \left(g^{(2)}_{\infty} - 1 \right) \mathrm{e}^{-t_{\text{adj}} / \tau_{\text{bunch}}} \right),
\]
which accounts for both antibunching and bunching behavior. In our case, since no bunching was observed and $g^{(2)}_{\infty} = 1$, the expression simplifies to:
\[
g^{(2)}(t) = 1 - (1 - g^{(2)}_0) \mathrm{e}^{-t_{\text{adj}} / \tau_{\text{antibunch}}},
\]
which was used to fit the experimental data. Here $t_{adj}$: adjusted time, representing the absolute value of the time delay. The fitting uses the Trust-Region Reflective method.

\noindent Moreover, in Fig. ~\ref{fig:noise_final1}a,b, the power-law fits are applied across both the decay and tail regions to quantify the rate of temporal decay and to reveal distinct dynamic signatures between the emitters.

\noindent For PSD, change-point detection was performed using the root mean square (RMS) statistic to identify significant transitions in the power spectral density data. The number of change points was limited to 2, and segments were determined by minimizing the RMS error within each segment.

\subsection{Training Attention-based Bi-LSTM Networks}

The attention-based bidirectional LSTM model was combined with Optuna for the following hyperparameter optimization:
\begin{itemize}
    \item Hidden layer size: Determines the number of neurons in each LSTM layer, affecting the model's capacity to learn complex patterns.
    \item Sequence length: Specifies the length of the input time window, allowing the model to consider temporal dependencies across a defined number of past time steps.
    \item Number of LSTM layers: Defines the depth of the network, with multiple layers enabling the model to learn higher-order temporal features.
    \item Dropout rate: Regularizes the network by randomly deactivating neurons during training, reducing the risk of overfitting.
    \item Learning rate: Controls the step size for updating the model’s weights during optimization, balancing convergence speed and stability.
\end{itemize}

\noindent The hyperparameter tuning process was driven by a validation loss criterion, ensuring the selection of parameters that minimized forecast error. The optimization procedure is formally described in Algorithm~\ref{alg:Attention_BiLSTM}, which outlines the end-to-end optimization of the attention-based Bidirectional LSTM (Bi-Attention-LSTM) model using Optuna. 

\noindent For each trial, different combinations of hyperparameters were sampled, and the dataset was divided into training (80\%), validation (10\%), and remotely isolated test sets (10\%), indicating future events. The model was trained using a bidirectional LSTM encoder, followed by an attention mechanism that dynamically weighs the encoded features according to their relevance to the prediction task. Early stopping was employed to prevent overfitting, ensuring the selection of the most effective model configuration. The model parameters were updated using the Adam optimizer, with the root mean squared error (RMSE) loss function guiding the learning process.

\noindent We utilized open-source Python libraries, including PyTorch for model development, Optuna for hyperparameter optimization, and Matplotlib for visualization. The model was trained on a dataset of 90 time steps using an Intel(R) Core(TM) i7-10700 CPU @ 2.90GHz. Each model took approximately 5 minutes to train for the full sequence length, with faster training times observed for shorter sequences.

\section{Data availability}
The data sets supporting the findings of this work are available from the corresponding authors upon reasonable request.
\section{Code Availability}
The code used for the order parameter and ML is available from the corresponding author on request.
\begin{algorithm}[H]
\caption{Attention-based Bi-LSTM Optimization with Optuna}
\begin{algorithmic}[1]
    \State \textbf{Require:} Normalized dataset $D_{norm}$, Original dataset $D_{orig}$
    \State \textbf{Require:} Hyperparameters $\{\eta, h, L, n, d\}$ (Learning rate, Hidden size, Sequence length, Layers, Dropout)
    \State \textbf{Require:} Maximum epochs $E_{\max}$, Trials $T$
    
    \For{$t \in [1, \dots, T]$} \Comment{Hyperparameter Optimization Loop}
        \State Sample hyperparameters: $\{\eta, h, L, n, d\} \sim p(\theta)$
        \State Split $D_{norm} \rightarrow \{D_{train}, D_{val}, D_{test}\}$
        \State Initialize model $\mathcal{M}_{\theta}$ with parameters $\theta$
        
        \For{$e \in [1, \dots, E_{\max}]$} \Comment{Training Loop}
            \State $x_{input} \rightarrow$ Bi-LSTM Encoder $\rightarrow H_e$
            \State $H_e, h_{dec} \rightarrow$ Attention Mechanism $\rightarrow C_e$
            \State $C_e \rightarrow$ Fully Connected Layer $\rightarrow \hat{y}_e$
            \State Compute RMSE Loss:
            \[
            \mathcal{L}_{\text{RMSE}} = \sqrt{ \frac{1}{N} \sum_{i=1}^{N} \left( y_i - \hat{y}_i \right)^2 }
            \]
            \State Update $\theta \leftarrow \theta - \eta \nabla_\theta \mathcal{L}_{\text{RMSE}}$
            
            \If{Early stopping criteria met}
                \State \textbf{Break}
            \EndIf
        \EndFor
        
        \State Compute validation loss $L_{val}$
        \State Generate denormalized predictions:
        \[
        \hat{y}_{denorm} = \hat{y}_e \cdot \sigma_{orig} + \mu_{orig}
        \]
        \State Save predictions and generate visualization $P_t$
    \EndFor

    \State \textbf{Output:} Optimized model $\mathcal{M}^*$ with minimum validation loss $L_{val}^*$
\end{algorithmic}
\label{alg:Attention_BiLSTM}
\end{algorithm}

\section{Acknowledgment}
The experimental part of this work was supported by the US Department of Energy, Office of Science, National Quantum Information Science Research Centers, Quantum Science Center. Low-temperature photoluminescence measurements were supported by the Center for Nanophase Materials Sciences (CNMS), which is a US Department of Energy, Office of Science User Facility at Oak Ridge National Laboratory. The theoretical and AI/ML parts of this research were supported by the QuPIDC, an Energy Frontier Research Center, funded by the US Department of Energy (DOE), Office of Science, Basic Energy Sciences (BES), under the award number DE-SC0025620. The authors gratefully acknowledge support with photolithography from Karthik Pagadala, as well as helpful discussions and valuable feedback from Dr. Demid Sychev and Dr. Alexander Senichev. 
\section{Authors contribution}
P.M. and A.V.K conceived this study; P.M.introduced the formalism with inputs from H.A. and A.V.K; B.L. and A.P. created the experimental setup; P.M. performed the experiments with some help from A.P. and B.L.; P.M. fabricated the Silicon Nitride quantum emitter sample; Y.C. and P.M. discussed the initial prediction algorithm and  Y.C. developed and implemented the final algorithm; S.B. implemented the code for the statistical analysis with the help of A.V.K and P.M.; P.M. wrote the manuscript with the help of A.V.K and H.A., and inputs from Y.C.; All authors discussed the results and contributed to writing the manuscript; A.B., V.M.S., and A.V.K. supervised the work.
\section{Competing interests}
The authors declare no competing interests.
\section{Additional Information}
\textbf{Supplementary information} The online version contains supplementary materials available at:
Supplementary information:
\begin{enumerate}
\item[S1.] Quasi-Quenched Regime and Trajectory-Resolved Analysis Enabled by Timescale Separation
\item[S2.] Comparison with Dynamic Decoupling and Optimal Control Methods
\item[S3.] Characterizing spectral diffusion decoherence channel in quantum emitters
\item[S4.] Order parameter estimation for the decoherence
\item[S5.] Relative stability of quantum system
\item[S6.] Theoretical Analysis of Anticipatory Framework for Spin Systems
\item[S7.] Sample preparation
\end{enumerate}

\clearpage
\bibliography{references}

\end{document}